\documentclass[twocolumn,prb,showpacs,preprintnumbers,dvips,floatfix]{revtex4}
\usepackage{graphicx}
\usepackage{dcolumn}
\usepackage{epsfig}

\newcommand{\bfr}{ {\bf r}}

\newcommand{\be}{\begin{equation}}
\newcommand{\ee}{\end{equation}}
\newcommand{\bea}{\begin{eqnarray}}
\newcommand{\eea}{\end{eqnarray}}

\newcommand\f{\ensuremath{\frac}}

\begin{document}
\title{Theory of electrostatically induced shape transitions in carbon nanotubes}
\author{Oleg E.\ Shklyaev, Eric Mockensturm }
\affiliation{Department of Mechanical and Nuclear Engineering, The Pennsylvania State University, University Park, PA, 16802-6300, USA}
\author{Vincent H.\ Crespi}
\affiliation{Department of Physics and Materials Research Institute, The Pennsylvania State University, University Park, PA, 16802-6300, USA}
\date{\today}

\begin{abstract}
A mechanically bistable single-walled carbon nanotube can act as a variable-shaped capacitor with a voltage-controlled transition between collapsed and inflated states. This external control parameter provides a means to tune the system so that collapsed and inflated states are degenerate, at which point the tube's susceptibility to diverse external stimuli-- temperature, voltage, trapped atoms -- diverges following a universal curve, yielding an exceptionally sensitive sensor or actuator that is characterized by a vanishing energy scale. For example, the boundary between collapsed and inflated states can shift hundreds of Angstroms in response to the presence or absence of a single gas atom in the core of the tube. Several potential nano-electromechanical devices can be based on this electrically tuned crossover between near-degenerate collapsed and inflated configurations.
\end{abstract}



 \maketitle

The equilibrium cross-sectional shape of a nanotube is controlled by a competition between elastic and surface energies. Sufficiently large-diameter tubes prefer the collapsed state, which captures the surface energy of the now-touching interior surfaces. Three distinct stability regimes can be defined in terms of the radius of the inflated, cylindrical state: below $R_1$ only the inflated state is stable; between $R_1$ and $R_2$ the inflated state remains stable, but the collapsed state is metastable; above $R_2$ collapse is stable and inflation is only metastable~\cite{Chopra95, Benedict98, Gao98, Tang05, Zhang06, Arash}. In the region of bistability above $R_1$, transitions between these two configurations can propagate down the axis of the tube~\cite{Chang08}. Here we show how the highly deformable conductive sp$^2$ sheet can act as a {\it non-linear, variable-shape capacitor} wherein electrostatic interactions within a charged tube shift $R_1$ and $R_2$ to favor the inflated state. A tube can be tuned by external voltage to a critical point at which the inflated and collapsed states are degenerate, producing a divergent susceptibility to diverse external stimuli and creating a regime of exceptionally sensitive nonlinear nano-electromechanical response.

\begin{figure}[!t]
\includegraphics[width=3.0in]{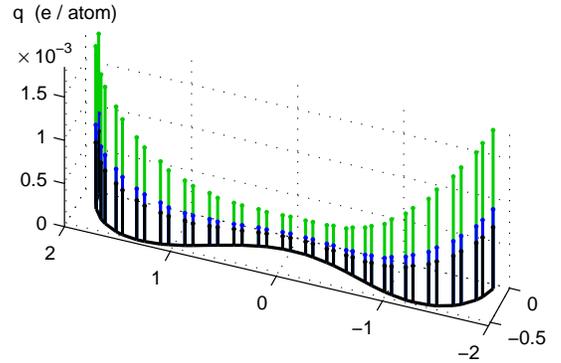}
\caption{Charge distribution over the collapsed configuration of (20,20) tube, showing the accumulation of charge in the bulbs. Dark blue and light green lines correspond to $U = 1$ and $U = 2$~V at $k_s = 0.1$~nm$^{-1}$, while black lines correspond to $k_s = 0.05$~nm$^{-1}$ at $U=1$~V.} \label{f:cr1}
\end{figure}

Consider an infinitely long single-walled tube of radius $R$ wherein an external voltage $U$ (relative to infinity) imparts additional charge $q$ onto the tube wall. We model the covalent interatomic interactions of this system with a Tersoff-Brenner potential~\cite{Brenner90,underestimate}, the non-bonded interactions with a Lennard-Jones term, and the electrostatic contribution via a screened Coulomb interaction and charging energy. The parameters $\epsilon$ and $\sigma$ of the Lennard-Jones interaction $\phi_{LJ}(r) = \epsilon \left( \f{\sigma}{r} \right)^{12} - 2 \epsilon \left( \f{\sigma}{r} \right)^6$ are $\sigma = 0.383$ nm and $\epsilon=2.39$ meV~\cite{Girifalco}. We exclude interatomic distances $r <  0.3$ nm from the Lennard-Jones term to protect the integrity of the covalent interactions. For computational efficiency, we neglect non-bonded interactions for $r > r_0$ nm and shift the Lennard-Jones potential upward by a small term linear in $r$ so that the energy and force vanish at the cutoff distance~\cite{r0, Zhang06}. In any real system, the induced charge on the tube wall is screened by countercharges in the  environment. In addition, coarse graining of the induced charge onto atoms requires an empirical on-site Coulomb self-interaction. Hence we introduce an on-atom self-energy that is linear in the induced charge and use a Thomas-Fermi screening at long distances:
\be
\phi_{\Sigma}(\bf r_i) = A q_i+\sum_{j\ne i}\f{q_j}{4\pi\epsilon_0}
\,\f{{\rm e}^{-k_sr_{ij}}}{r_{ij}}. \label{eq:distrib}
\ee
The empirical parameter $A = 7.8$ nm$^{-1}$ is fitted to results when $k_s=0$~\cite{Keblinski02}, but the precise value of $A$ has only minor effects on our main results. Thomas-Fermi screening provides a reasonably accurate model of screening due to e.g.\ an electrolyte. In $10^{-1}$ to $10^{-4}$ M NaCl solutions the Debye screening length varies from 30 to 1 nm~\cite{Lyklema}; we use $k_s=0.2, 0.1, 0.05$ nm$^{-1}$. The net electrostatic self-interaction is $E_{\Sigma} = \f{1}{2} \sum_{i=1}^{ N}q_i\phi_{\Sigma} (r_{i})$. At fixed applied voltage $U$, the tube when collapsed holds less charge than when inflated, yielding a contribution $E_Q= -qU$ due to work against this external voltage. The total energy is then a sum of covalent ($E_{TB}$), non-bonded ($E_{LJ}$), and electrostatic ($E_{\Sigma} + E_Q$) contributions. The energy differences $\Delta E$ quoted below are always relative to the energy of the corresponding inflated state at the same voltage, so that $\Delta E = 0$ corresponds to degeneracy. Equations of motion (under periodic boundary conditions) are integrated with a Verlet algorithm, incorporating viscous damping to relax to static equilibria.

A constant-voltage condition models a tube connected to an electrode with a quasi-ohmic contact; (a constant-charge condition could model a tunneling contact, with a distinct type of collapse/inflation transition dynamics described later). When the tube is collapsed, the charge accumulates in the bulbs, as shown in Fig. \ref{f:cr1}. To find the charge distribution, we write the potential $\phi_{\Sigma}(\bfr_i)$ given by (\ref{eq:distrib}) for each atom $1\le i\le N$ in the unit cell and solve the system of self-consistent linear equations:
\begin{equation}
\begin{array}{c}
\displaystyle     \sum^{j=N}_{j\ne i}q_j\sum_{n=-\infty}^{n=\infty}\f{{\rm e}^{-k_s\rho_{ijn}}}{\rho_{ijn}}\;+ \qquad \qquad \qquad \qquad \qquad \qquad\\
\displaystyle  \qquad
q_i\bigg(A+{\sum_{n=-\infty}^{n=\infty}}^{\prime}\f{{\rm
e}^{-k_s|nL|}}{|nL|} \bigg) =4\pi\epsilon_0\phi_{\Sigma}(\bfr_i),
\end{array}\label{eq:distrib1}
\end{equation}
with respect to unknown charges $q_i$ at a fixed electrostatic potential $\phi_{\Sigma}(\bfr_i)=U$, where $\rho_{ijn}$ is the distance between atoms $i$ and $j$ separated by $n$ unit cells. The primed sum omits $n=0$.

\begin{figure}[!t]
\includegraphics[width=3.2in]{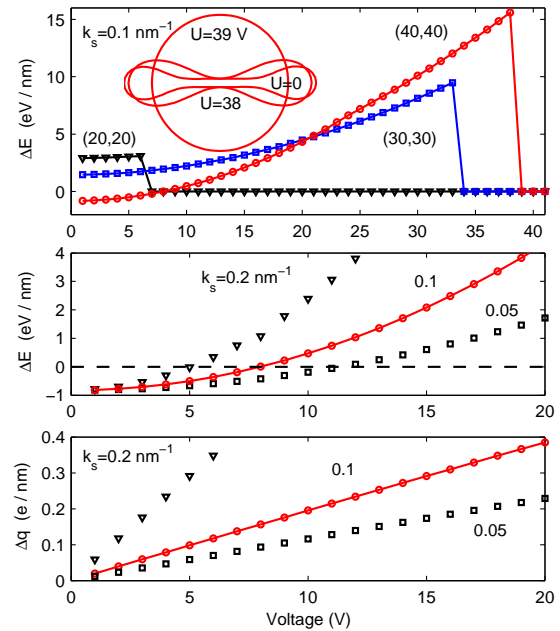}
\caption{Upper panel: Net energy $\Delta E=\Delta E_{TB} + \Delta E_{LJ} + \Delta E_{\Sigma} + \Delta E_Q$ per unit length of (20,20), (30,30) and (40,40) tubes as a function of the applied voltage $U$, where $\Delta E = 0$ corresponds to the inflated state at the same voltage. Discontinuities correspond to abrupt transitions into the inflated state. The inset shows changes of the shape of a (40,40) tube under external voltage. Middle panel: $\Delta E$ for a (40,40) tube when $k_s = 0.2, 0.1, 0.05 \mathrm{nm}^{-1}$. Lower panel: charge difference between the inflated and collapsed configurations for a (40,40) tube; the near-linearity indicates that the end-state tube shapes are relatively insensitive to charge (i.e. the capacitances of inflated and collapsed states are each roughly constant across this range of voltage).}
\label{f:ne1}
\end{figure}

Fig.~\ref{f:cr1} shows the excess charge per atom. As regions of high curvature, the bulbs accumulate charge, more so at higher voltages. More weakly screened charge distributions require less total charge to sustain a given voltage. For strong screening, the on-site self-energy in Eqn. (\ref{eq:distrib1}) is more important compared to the inter-atomic interaction, so the system becomes slightly more sensitive to the empirical parameter $A$~\cite{screen}.

The inflated state holds more charge than the collapsed state at fixed voltage, therefore it has a higher capacitance and is favored under increasing charge. The upper panel of Fig.~\ref{f:ne1} depicts voltage-controlled shape transitions for (20,20), (30,30) and (40,40) tubes. All three systems are initialized to a collapsed state at $U=0$; (collapse  is metastable for (20,20) or (30,30) and stable for (40,40)). Charging not only favors the higher-capacitance inflated state, but also decreases the barrier against inflation. At $U_{\mathrm{crit}}(R)$ the kinetic barrier against inflation disappears and the tube inflates. The inset in the figure depicts this transition for the (40,40) tube. As the voltage increases, the bulbs expand and the flattened interior shrinks. At $U_{\mathrm{crit}}\approx39$~V the tube snaps open. The middle and lower panels of Fig.~\ref{f:ne1} demonstrate influence of the screening parameter $k_s$ on the energy $\Delta E$ and charge difference $\Delta q$ correspondingly. The effect is robust across a wide range of screening lengths. Of particular interest are the more modest voltages needed to tune the system to the degeneracy point $\Delta E = 0$, a condition in which the system becomes exceptionally sensitive to external perturbations, as described below.

\begin{figure}[!t]
\includegraphics[width=3.2in]{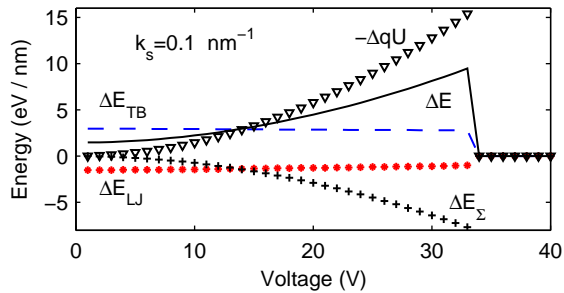}
\caption{Decomposition of the energy difference $\Delta E$ between collapsed and inflated states for a charged (30,30) tube into covalent/elastic ($E_{TB}$), Lennard-Jones ($E_{LJ}$), Coulomb ($E_{\Sigma}$), and charging ($\Delta qU$) contributions, all measured per unit length for $k_s = 0.1 \mathrm{nm}^{-1}$.} \label{f:34a}
\end{figure}

Fig.~\ref{f:34a} depicts the various contributions to $\Delta E$ (solid line), the energy difference between collapsed and inflated states, as functions of $U$ for the (30,30) tube. The elastic contribution (blue dashed line) is always positive, since a uniform circular cross-section minimizes the total curvature energy. Conversely, the Lennard-Jones contribution (red dots) is always negative, since the collapsed state allows closer approach of opposing surfaces. Since the collapsed state has a lower capacitance, the Coulomb self-energy $\Delta E_{\Sigma}$ (pluses) is negative (at fixed voltage). The charging contribution $\Delta E_Q$ (triangles) associated with motion of excess charge on or off the tube is positive, due to the same difference in capacitance.

\begin{figure}[!t]
\includegraphics[width=2.0in]{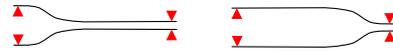}
\caption{Schematic depicting the evolution of a double-pinned tube from a mostly collapsed state at $U<U_{\mathrm{degen}}$ to a mostly inflated state at $U>U_{\mathrm{degen}}$.} \label{f:vol}
\end{figure}

For tubes with $R > R_2$, like (40,40), the external voltage can be tuned so that collapsed and inflated states are degenerate, i.e. $\Delta E = 0$. Since this $U_{\mathrm{degen}} < U_{\mathrm{crit}}$, the two states are separated by a barrier. If boundary conditions that eliminate this barrier can be imposed, then the susceptibility against perturbations that favor collapse or inflation will {\it diverge} at this critical voltage. For example, a tube with $R > R_2$ held inflated at one end and collapsed at the other (as shown in Fig.~\ref{f:vol}) must contain a transition region. The kinetic barrier between degenerate configurations is thereby eliminated, since the transition region can move freely along the tube axis. Such pinning has already been produced experimentally, even in the first discovery paper on nanotube collapse~\cite{Chopra95}, since the rigid end-cap of a tube holds the end of the tube open even after the interior collapses. This pinned-open/pinned-closed configuration provides many possible modes of device operation; these can be classified as those that operate {\it around} the degeneracy point versus those that operate {\it at} the degeneracy point.

\vspace{1 mm} \noindent {\bf Devices that operate {\it around} the degeneracy point:} By sweeping the applied voltage across $U_{\mathrm{degen}}$, a doubly pinned nanotube can act as a electrical-mechanical transducer, transitioning between the two states shown in Fig.~\ref{f:vol}. In actuator mode, the inflating tube could perform work over long axial distances (by pushing a load along the axis) or over short transverse distances (limited by the diameter of the inflated state). Axial motions could couple to either liquids inside the tube or solids that are attached to the tube exterior. Reversing the transduction, the system could also convert mechanical motion into charge, similar to a piezoelectric sensor. Since the collapsed state has the smaller capacitance, a charge $\Delta q$ will leave the tube when an external compressive load collapses the tube. For example, approximately 200~eV of work will collapse a 100~nm length of a (40,40) tube held at $U = 15$~V (for $k_s=0.1$), producing a charge signal of $\Delta q \sim 30e$, (Fig.~\ref{f:ne1}) well within the range of modern measurement techniques~\cite{Bylander, Keller, Grabert}. Unlike piezoelectric crystals whose fractional capacity to elongate $\Delta l/l$ is at most on the order of 10$^{-3}$, bistable nanotube devices -- for axial transport -- can have a range of motion comparable to the length of the device itself. The efficiency of this transducer varies considerably depending on the operating voltage, but is comparable to that of established piezoelectric actuators~\cite{Fleming} across a broad range of operation.

The time response of these nonlinear nano-mechanical systems is governed by the axial speed of the transition front between collapsed and inflated states. This front moves at $v \propto \sqrt{\Delta E}$~\cite{Chang08}; in our system $\Delta E$ is proportional to the voltage deviation from $U_{\mathrm{degen}}$. For example, a ten-volt swing away from $U_{\mathrm{degen}}$ for a (40,40) tube yields $\Delta \mathrm{E}_a \sim 2$~meV per carbon atom; taking $m_C$ as a characteristic mass, one obtains $v \sim \sqrt{\Delta \mathrm{E}_a/m_C} \sim 200$ m/s. A typical device dimension of 0.1 micron then implies roughly GHz operating frequencies.


\begin{figure}[!t]
\includegraphics[width=3.3in]{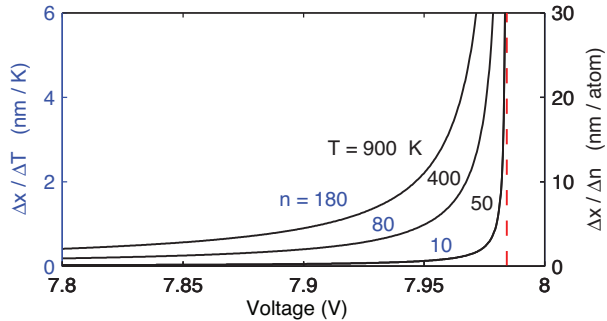}
\caption{Universal response function to temperature $\Delta x/\Delta T$ (or atom number $\Delta x/\Delta n$) for different number of encapsulated atoms $n$ (or temperature $T$). The red dashed line marks $U_{\rm degen}$.} \label{f:respa}
\end{figure}

\vspace{1 mm} \noindent {\bf Devices that operate {\it at} the degeneracy point:} Of particular interest are devices designed to operate as close to degeneracy as possible. Such a system is exceptionally sensitive to external perturbations that upset the balance between inflated and collapsed states. For example, a gas trapped in the interior adds a $PV$ term to the free energy, so that equilibrium is obtained when the gas pressure balances the energy density associated with the volume change between collapsed and inflated states: $P = \Delta E /\Delta A$ where $\Delta A$ is a change in the cross section area and $\Delta E$ is measured per unit axial length; (this simple analysis neglects fluctuations, which are discussed later). For an ideal gas, one obtains the differential sensitivity of the length $x$ of the inflated portion to changes in either atom number $n$ or temperature $T$: 
\be 
\label{eq:response0} \displaystyle \Delta x= -\f{nk_{B}\Delta T}{\Delta E} = -\f{k_{B}T\Delta n }{\Delta E}. 
\ee 
Tuning the voltage to $U_{\mathrm{degen}}$, we obtain $\Delta x \rightarrow \infty$. Even a single interior gas atom could shift the transition zone by a substantial amount, limited by the precision with which the system energetics can be tuned. Expanding $\Delta E(U)$ around $U=U_{\mathrm{degen}}$ for a (40,40) tube with $k_s=0.1$, we can obtain the response functions to temperature ($\Delta x/ \Delta T$) or atom number ($\Delta x/ \Delta n$) as a universal family of curves plotted in Fig.~\ref{f:respa}. These curves are interpretable as either the sensitivity to temperature at fixed number of interior gas atoms or the sensitivity to atom number at fixed temperature. Other external perturbations can also elicit strong responses: for example, dramatic responses can be anticipated to surface acoustic wave excitation, a highly nonlinear version of the nanotube charge pumping that was previously examined~\cite{Talyanskii} theoretically and experimentally for regular inflated tubes. Adsorbates and variations in temperature~\cite{Chang10} or mechanical boundary conditions provide additional experimental handles to tilt the delicate balance between configurational states.

Collapsed-based nano-electromechanical devices could also be operated in constant-charge mode rather than constant-voltage mode, a distinction similar to that between constant-pressure and constant-volume. At constant charge the $qU$ term is absent but the inflated state continues to be favored at higher charge per unit length. Around the degeneracy point the system self-organizes into a mixed collapsed/inflated state, with higher charge density in the inflated section than in the collapsed one. This charge bubble -- a giant polaron of sorts -- could be subject to further experimental manipulation through scanned probes or split gates.

Although huge, tunable responses to perturbations are possible in this system, non-uniformities along the tube will prevent perfect tuning and hence cut off the divergent susceptibility. For example, voltage fluctuations of $\delta U=10$~mV are characteristic of nanotubes or graphene on silica substrates~\cite{graphene1} (although more uniform environments are possible on alternative substrates~\cite{graphene2}). An inhomogeneity of $\delta U=10$~mV cuts off $dX/dn$ at $\sim$ 18~nm/atom at $T=400$ K. Thermal fluctuations should be particular important in this nanoscale, one-dimensional system. A simple one-degree-of-freedom model (which assumes a constant shape to the transition region) writes the energy of the system as $\Delta Ex$ where $x$ is the location of the collapse-inflation transition. Simple thermodynamics then implies that $\sqrt{\langle x^2\rangle-\langle x\rangle^2} = \f{kT}{\Delta E}$. Since the system has only one characteristic length scale, the same universal divergence results, with fluctuations in the location of the collapse/inflation transition of the same size as the increment in location induced by a single additional interior gas atom (Eqn. \ref{eq:response0}b). This distance can be many nanometers and could be exploited in fluctuation-based device modalities similar to those of biological molecular motors.


We thank Paul Lammert and Ilya Grigorenko for valuable discussions and acknowledge NSF CMMI-0727890 and DMR-0707332 for support.


\begin{references}

\bibitem{Chopra95}
N.\ G.\  Chopra, L.\ X.\ Benedict, V.\ H.\ Crespi, M.\ L.\ Cohen, S.\ G.\ Louie, A.\ Zettl, Nature {\bf 377}, 135 (1995).

\bibitem{Benedict98}
L.\ X.\ Benedict, V.\ H.\ Crespi, N.\ G.\ Chopra, A.\ Zettl, M.\ L.\ Cohen, S.\ G.\ Louie, Chem. Phys. Lett.  {\bf 286}, 490 (1998).

\bibitem{Gao98}
G.\ Gao, T.\ Cagin, W.\ A.\ Goddard, Nanotechnology {\bf 9}, 184 (1998).

\bibitem{Tang05}
T.\ Tang, A.\ Jagota, C.-Y.\ Hui,  N.\ J.\ Glassmaker, J. Appl. Phys. {\bf 97}, 074310 (2005).

\bibitem{Zhang06}
S.\ Zhang, R.\ Khare,  T.\ Belytschko, K.\ J.\ Hsia, S.\ L.\ Mielke, G.\ C.\ Schatz, Phys. Rev. B {\bf 73}, 075423 (2006).

\bibitem{Arash}
E.\ Mockensturm, A.\ Mahdavi, V.\ Crespi, Proc. of IMECE 2005, 82991 (2005).

\bibitem{r0}
Following\protect\cite{Zhang06}, the cutoff distance $r_0$ was chosen close to 1~nm.

\bibitem{Chang08}
T.\ Chang, Phys. Rev. Lett. {\bf 101}, 175501 (2008).

\bibitem{Brenner90}
D.\ W.\ Brenner, Phys. Rev. B {\bf 42}, 9458 (1990).

\bibitem{underestimate}
The Tersoff-Brenner potential underestimates the bending stiffness of an sp$^2$ sheet: writing the strain energy of a nanotube as $D'/(2R)^2$,~\cite{Tibbetts83} one obtains $D' = 0.08, 0.046$ eV nm$^2$/atom for first-principles density functional theory~\cite{Sanchez99, Yakobson01}, and the first generation~\cite{Robertson92, Yakobson01} Tersoff-Brenner potentials, respectively. To simplify numerical simulations, we adopt the first-generation potential (the capabilities of the second-generation potential regarding close atomic approaches are not needed here). This underestimation of the bending stiffness does not change any of our main conclusions, but it does shift all of the stability regimes towards smaller radii.

\bibitem{Girifalco}
L.\ A.\ Girifalco, M.\ Hodak, R.\ S.\ Lee,  Phys. Rev. B {\bf 62}, 13104 (2000).

\bibitem{Keblinski02}
P.\ Keblinski, S.\ K.\ Nayak, P.\ Zapol, P.\ M.\ Ajayan, Phys. Rev. Lett. {\bf 89}, 255503 (2002).

\bibitem{Lyklema}
J.\ Lyklema (1993).  Fundamentals of interface and colloid science. Academic press, NY.

\bibitem{screen}
In computations performed for (20,20) tube at $U=5$~V, a ten percent variance in $A=7.8$ produced less than 1\% errors in relative tube energy, $\Delta E$, for $k_s=0.1, 0.05, 0.025$~nm$^{-1}$.

\bibitem{Tibbetts83}
G.\ G.\ Tibbetts, J. Crys. Growth {\bf 66}, 632 (1983).

\bibitem{Sanchez99}
D.\ Sanchez-Portal, E.\ Artacho, J.\ M.\ Soler, A.\ Rubio, P.\ Ordejon, Phys. Rev. B 59, 12678 (1999)

\bibitem{Robertson92}
D.\ H.\ Robertson, D.\ W.\ Brenner, J.\ W.\ Mintmire, Phys. Rev. B {\bf 45}, 12592 (1992)


\bibitem{Yakobson01}
B.\ I.\ Yakobson, P.\ Avouris, Carbon Nanotubes {\bf 80}, 287 (2001).

\bibitem{Brenner02}
D.\ W.\ Brenner, O.\ A.\ Shenderova, J.\ A.\ Harrison, S.\ J.\ Stuart, B.\ Ni, S.\ B.\ Sinnott, J. Phys.: Condens. Matter {\bf 14} 783 (2002).

\bibitem{Garcia07}
D.\ Garcia-Sanchez, A.\ San Paulo, M.\ J.\ Esplandiu, F.\
Perez-Murano, L.\ Forro, A.\ Aguasca,  A.\ Bachtold, Phys. Rev.
Lett. {\bf 99}, 085501 (2007).

\bibitem{Bylander}
J.\ Bylander, T.\ Duty, P.\ Delsing, Nature, {\bf 434}, 361, (2005).

\bibitem{Keller}
M.\ W.\ Keller, J.\ M.\ Martins, N.\ M.\ Zimmerman, A.\ H.\ Steinbach, Appl. Phys. Lett. {\bf 69} (12), 1804, (1996).

\bibitem{Grabert}
H.\ Grabert, M.\ H.\ Devoret, Single Charge Tunneling: Coulomb Blockade Phenomena in Nanostructures (Plenum, New York, 1992)

\bibitem{Fleming}
S.\ O.\ Moheimani, A.\ J.\ Fleming, (2006) Piezoelectric transducers for vibration control and damping. Springer.

\bibitem{Talyanskii}
V.\ I.\ Talyanskii, D.\ S.\ Novikov, B.\ D.\ Simons, L.\ S.\ Levitov,  Phys. Rev. Lett. {\bf 87} (27), 276802, (2001); V.\ I.\ Talyanskii, P.\ Leek, M.\ Buitelaar, C.\ G.\ Smith, D.\ Anderson, J.\ Jones, J.\ Wei, D.\ Cobden, Physica E {\bf 34}, 662, (2006).

\bibitem{Chang10}
T.\ Chang, Z.\ Guo, Nano Lett. {\bf 10}, 3490 (2010).

\bibitem{graphene1}
J.\ Martin, N.\ Akerman, G.\ Ulbright, T.\ Lohmann, J.\ H.\ Smet, K.\ von Klitzing, A.\ Yacoby, Nature, Physics, {\bf 4}, 144 (2008).

\bibitem{graphene2}
X.\ Hong, A.\ Posadas, K.\ Zou, C.\ H.\ Ahn, J.\ Zhu, Phys. Rev. Lett. {\bf 102}, 136308 (2009).

\end{references}
\end{document}